\documentclass{aa}  
\pdfoutput=1
\usepackage[varg]{txfonts} 
\usepackage{natbib}
\usepackage{graphicx}
\usepackage{amsmath}	
\usepackage{amssymb}	
\usepackage{hyperref}
\hypersetup{
    colorlinks   = true, 
    urlcolor     = blue, 
    linkcolor    = blue, 
    citecolor   = blue 
}
\usepackage{multirow}
\usepackage{tabularx}
\usepackage[group-separator={,}]{siunitx}
\usepackage{bm}
\usepackage{xcolor}
\usepackage[export]{adjustbox}
\usepackage{url}

\usepackage[rightcaption]{sidecap}

\defcitealias{KV450_2020AA...633A..69H}{H20}
\newcommand{\Htwo}{\citetalias{KV450_2020AA...633A..69H}}
\defcitealias{ImSim_2019AA...624A..92K}{K19}
\newcommand{\Kone}{\citetalias{ImSim_2019AA...624A..92K}}
\defcitealias{KV450_2019AA...632A..34W}{W19}
\newcommand{\Wone}{\citetalias{KV450_2019AA...632A..34W}}

\usepackage[normalem]{ulem}

\begin{document} 

   \title{KiDS+VIKING-450: An internal-consistency test for cosmic shear tomography with a colour-based split of source galaxies}
   \titlerunning{KiDS+VIKING-450: An internal-consistency test for cosmic shear tomography}
   \authorrunning{S.-S. Li et al.}

\author{Shun-Sheng Li\inst{1}
        \and
        Konrad Kuijken\inst{1}
        \and
        Henk Hoekstra\inst{1}
        \and\\
        Hendrik Hildebrandt\inst{2}
        \and
        Benjamin Joachimi\inst{3}
        \and
        Arun Kannawadi\inst{4}
}

\institute{Leiden Observatory, Leiden University, Niels Bohrweg 2, 2333 CA Leiden, the Netherlands\\
            \email{ssli@strw.leidenuniv.nl}
            \and
            Ruhr-Universit\"at Bochum, Astronomisches Institut, German Centre for Cosmological Lensing (GCCL), Universit\"atsstr. 150, 44801 Bochum, Germany
            \and
            Department of Physics and Astronomy, University College London, Gower Street, London WC1E 6BT, UK
            \and
            Department of Astrophysical Sciences, Princeton University, 4 Ivy Lane, Princeton, NJ 08544, USA
             }

   \date{Received 25 August 2020 / Accepted 21 December 2020}

\abstract
{
We performed an internal-consistency test of the KiDS+VIKING-450 (KV450) cosmic shear analysis with a colour-based split of source galaxies. Utilising the same measurements and calibrations for both sub-samples, we inspected the characteristics of the shear measurements and the performance of the calibration pipelines. On the modelling side, we examined the observational nuisance parameters, specifically those for the redshift calibration and intrinsic alignments, using a Bayesian analysis with dedicated test parameters. We verified that the current nuisance parameters are sufficient for the KV450 data to capture residual systematics, with slight deviations seen in the second and the third redshift tomographic bins. Our test also showcases the degeneracy between the inferred amplitude of intrinsic alignments and the redshift uncertainties in low redshift tomographic bins. The test is rather insensitive to the background cosmology and, therefore, can be implemented before any cosmological inference is made.
}

\maketitle
%

\section{Introduction}\label{Sec:intro}

Cosmic shear, the coherent distortion of distant galaxy shapes that arises from weak gravitational lensing by large-scale structures, is sensitive to the amplitude of matter density fluctuations, which are usually quantified by $\sigma_8$\footnote{The standard deviation of linear-theory density fluctuations in a sphere of radius $8h^{-1}~{\rm Mpc}$, where $H_0=100h~{\rm km~s^{-1}~Mpc^{-1}}$.} and to the mean matter density $\Omega_m$. Therefore, the main result from a cosmic shear survey is conventionally reported as a derived parameter $S_8\equiv \sigma_8\left(\Omega_{\rm m}/0.3\right)^{0.5}$. Alternatively, the cosmic microwave background (CMB) measurements can infer the local density fluctuations by extrapolating the measured amplitude of temperature fluctuations at recombination, assuming a cosmological model. Hence, by comparing the results from these two different probes, we can test the cosmological models.

As for the current standard model of cosmology, dubbed $\Lambda$ cold dark matter ($\Lambda$CDM), the latest results from \emph{Planck} ~\citep{Planck_2018arXiv180706209P} yield a constraint of $S_8=0.832\pm0.013$ ($68\%$ credible region), which is slightly higher than the results from the recent cosmic shear surveys, such as the Dark Energy Survey (DES; \citealt{DES_2018PhRvD..98d3528T}, $S_8=0.782^{+0.027}_{-0.027}$), the Hyper Suprime-Cam Subaru Strategic Program (HSC; \citealt{HSC_2019PASJ...71...43H}, $S_8=0.780^{+0.030}_{-0.033}$), and especially the Kilo-Degree Survey (KiDS; \citealt{KV450_2020AA...633A..69H}, hereafter H20, $S_8=0.737^{+0.040}_{-0.036}$). 

In the era of `precision cosmology’, we have to be careful about any potential systematic effects associated with observations when interpreting results from different surveys. Given this consideration, performing internal-consistency checks is a standard part of any cosmological probe. A cosmic shear study typically bases its internal-consistency tests on a split of the estimated two-point shear correlations~(\citealt{consistency_2019MNRAS.484.3126K}; or Sect.~7.4 of \Htwo). By assigning duplicated model parameters to each subset, one can perform theoretical modelling of the reconstructed data vector and quantify the data consistency by comparing the duplicated model parameters. This approach is useful to check for potential inconsistencies for a specific sample of source galaxies. However, the robustness is only tested at a late stage of the analysis, whilst doubling cosmological parameters comes at a considerable computational cost. The latter prevents further splits of the source sample in practice, whereas such splits can be particularly interesting because the systematics may differ.

Source galaxy properties challenge the calibration pipelines in mainly two ways: the shape measurements and the redshift estimates. First, different galaxy samples usually have different distributions of ellipticities, with red, early-type galaxies tending to have rounder shapes than their blue, late-type counterparts~(\citealt{shape_2019ApJ...871...76H,ImSim_2019AA...624A..92K}, hereafter K19). This introduces a correlation between the shear bias and underlying galaxy sample, mainly because the shape measurements are sensitive to the distributions of galaxy ellipticities, for example, the \emph{lens}fit algorithm used in the KiDS survey assigns weights to the measured ellipticities, resulting in a bias towards intermediate ellipticity values~\citep{ImSim_2017MNRAS.467.1627F}. Second, both the accuracy and the precision of a photometric redshift estimate depends on broad spectral features of a galaxy, for example, the Balmer break below $4000\AA$~\citep{review_2019NatAs...3..212S}. The significance of these broad spectral features varies by galaxy spectral types. Generally speaking, galaxies with an old stellar population appear red at rest-frame optical wavelengths and have a pronounced $4000\AA$ break. The bluer the galaxy, the more young stars it contains, washing out the Balmer break and other broad spectral features. Therefore, the error in photometric redshifts correlates with the galaxy spectral type~\citep{book_2010gfe..book.....M}.

We consider these sample-related systematic effects, specifically the photometric redshift uncertainty, in the KiDS cosmic shear analysis. We split the source galaxies into two mutually exclusive sub-samples according to their spectral types and apply the same measurement and calibration pipelines to these two sub-samples. This way we explored how sample-related systematics can alter the measurements and how well the calibration pipelines can assuage these effects. This split also has implications for the modelling of intrinsic alignments, which have to be taken into account explicitly. To quantify the consistency, we performed a Bayesian analysis with dedicated test parameters describing relative deviations of the nuisance parameters between the two sub-samples. By checking their posterior distributions, we can verify if the original setting suffices to capture the residual biases. The analysis code is publicly available\footnote{\url{https://github.com/lshuns/CosmicShearRB}}.

Our approach complements other studies that check for consistency in the inferred cosmological parameters by removing tomographic bins~\citep{consistency_2019MNRAS.484.3126K}, or by splitting the sample by galaxy type \citep{IA_2019MNRAS.489.5453S}, whilst marginalising over the corresponding nuisance parameters. We explored a different aspect: we fixed cosmological parameters but explored changes in the nuisance parameters instead. We found that our approach can test for inconsistencies in the redshift distributions and highlights the degeneracy between the redshift uncertainties and the apparent intrinsic alignment signals in a cosmology-insensitive fashion.

This paper is organised as follows. In Sect.~\ref{Sec:data}, we briefly describe the cosmic shear catalogues under consideration. We show the redshift calibration in Sect.~\ref{Sec:cal_z} and the shear bias calibration in Sect.~\ref{Sec:cal_shear}. We then introduce measuring and modelling the shear signal in Sect.~\ref{Sec:shear}. We discuss the covariance matrix and the consistency tests in Sect.~\ref{Sec:test}. The main results are presented in Sect.~\ref{Sec:resul}, and we summarise in Sect.~\ref{Sec:discussion}.

\section{Data}\label{Sec:data}

Our test is based on the first release of optical+infrared KiDS cosmic shear data dubbed KiDS+VIKING-450 (KV450; \citealt{KV450_2019AA...632A..34W}, hereafter W19)\footnote{\url{http://kids.strw.leidenuniv.nl/DR3/kv450data.php}}. It includes four-band optical photometry ($ugri$) from the first three data releases of KiDS~\citep{KiDSdata_2015AA...582A..62D,KiDSdata_2017AA...604A.134D} and five-band near-infrared photometry ($ZYJHK_{\rm s}$) from the overlapping VISTA Kilo-Degree Infrared Galaxy Survey (VIKING, \citealt{VIKING_2013Msngr.154...32E}).

Details on the derivation and verification of these cosmic shear catalogues can be found in the main KiDS cosmic shear papers~(\citealt{K450_2017MNRAS.465.1454H}; \Htwo) and their companion papers~(\citealt{ImSim_2017MNRAS.467.1627F}; \Wone). For reference, the public catalogues contain all of the necessary information to conduct a tomographic cosmic shear analysis. Amongst the most important columns are the photometric redshifts (photo-$z$, or $z_{\rm B}$ as in the catalogues) and the galaxy shapes (described by two ellipticity components $\epsilon_1$, $\epsilon_2$). The $z_{\rm B}$ values are estimated using the Bayesian photometric redshift code~\citep[BPZ;][]{BPZ_2000ApJ,BPZ_2006AJ} with an improved redshift prior from \citet{BPZprior_2014ApJ} and the nine-band photometry from \Wone. The galaxy shapes are measured from the $r$-band images (median seeing $0.7''$) using the \emph{lens}fit algorithm~\citep{lensfit_2007MNRAS.382..315M,lensfit_2008MNRAS.390..149K,lensfit_2013MNRAS.429.2858M} with a self-calibration for noise bias~\citep{ImSim_2017MNRAS.467.1627F}.

\begin{figure}
   \centering
   \includegraphics[width=\hsize]{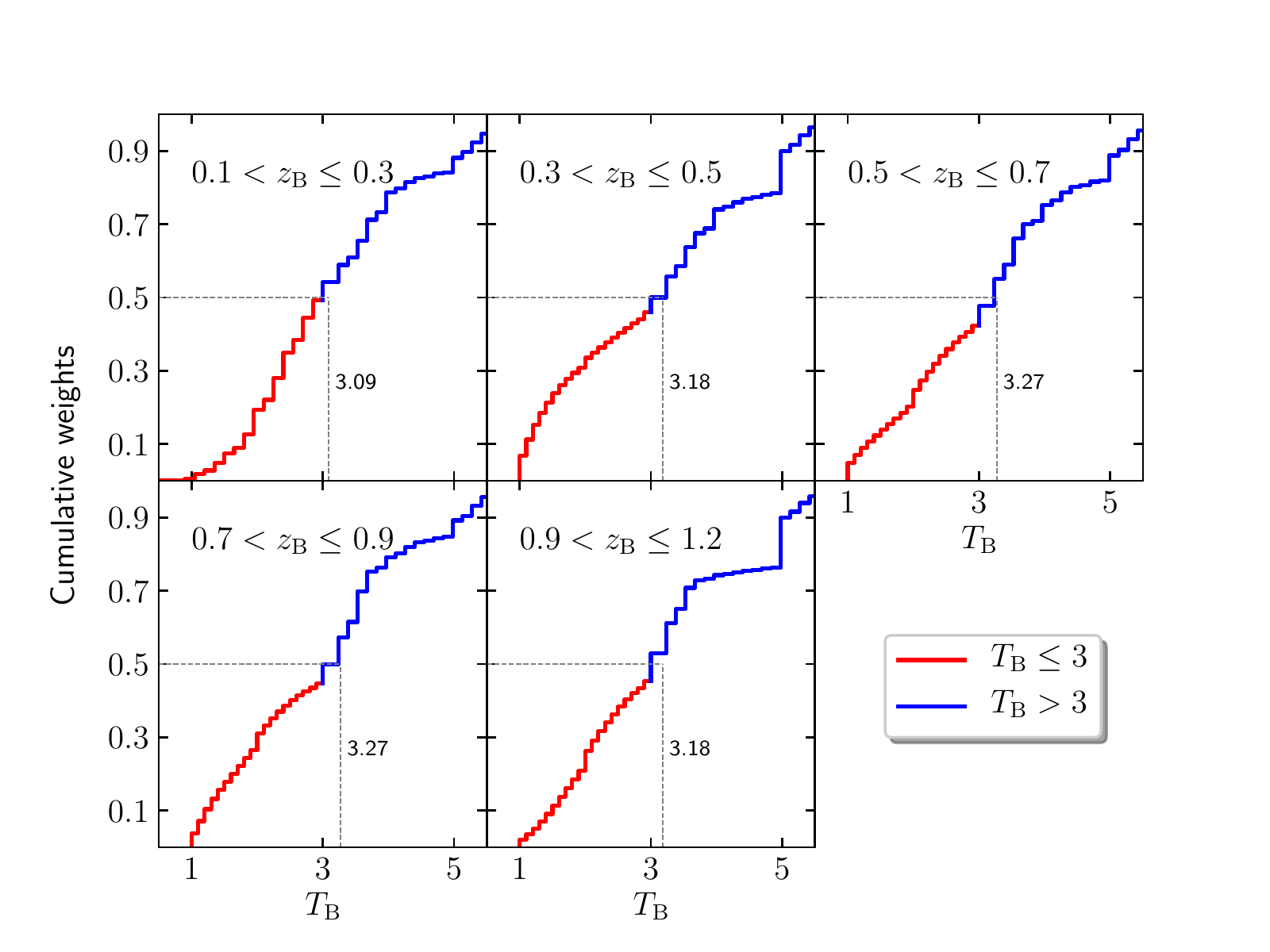}
      \caption{Cumulative \emph{lens}fit-weighted distributions of $T_{\rm B}$ values. The dashed line indicates the ideal half-half split in each tomographic bin, which is close to our split at $T_{\rm B}=3$.}
         \label{fig:TB}
   \end{figure}

Throughout this study, we only used sources with valid nine-band photometry ({\tt GAAP\_Flag\_ugriZYJHKs}==0). This mask reduces the original area by $\sim 5\%$ and retains $\sim 13$ million objects, which is identical to the choice made by the main KV450 cosmic shear analysis. Following \Htwo, we binned source galaxies into five tomographic bins defined as $0.1<z_{\rm B}\leq 0.3$, $0.3<z_{\rm B}\leq 0.5$, $0.5<z_{\rm B}\leq 0.7$, $0.7<z_{\rm B}\leq 0.9$, $0.9<z_{\rm B}\leq 1.2$. Given the purposes of checking systematic effects caused by galaxy properties, we further split the whole sample into two sub-samples according to the spectral types of source galaxies. This is achieved by using the $T_{\rm B}$ values reported by the \textsc{BPZ} code during the photo-$z$ estimating procedure~\citep[see][for a detailed discussion]{BPZ_2000ApJ}. Briefly, the $T_{\rm B}$ value is calculated within a Bayesian framework using six templates of galaxy spectra~\citep{BPZT_1980ApJS...43..393C,BPZT_1996ApJ...467...38K}. We defined our two sub-samples as $T_{\rm B}\leq 3$ (a combination of E1, Sbc, Scd types, labelled as `red' in this paper) and $T_{\rm B}>3$ (a combination of Im and two starbust types, labelled as `blue' in this paper). This cut is chosen to ensure similar statistical power in the two sub-samples (see Fig.~\ref{fig:TB}). Source properties of these two sub-samples are summarised in Table~\ref{Table:basic}.

%

%
\begin{table*}
\caption{\label{Table:basic}Source information in the two sub-samples.}
\centering
\begin{tabular}{lccrccrrr}
\hline\hline
Sample & Bin & Photo-$z$ range & Total \emph{lens}fit weights & $n_{\rm eff}$ & $\sigma_{\epsilon,i}$ & $m$-bias & Mean$\left(z_{\rm DIR}\right)$ & Median$\left(z_{\rm DIR}\right)$ \\
 & & & & $\left({\rm arcmin}^{-2}\right)$ & & & \\

\hline
$T_{\rm B}\leq 3$ & 1 & $0.1< z_{\rm B} \leq 0.3$ & \num{7031963} & $0.38$ & $0.279$ & $-0.029\pm 0.010$ & $0.351$ & $0.282$ \\
(red)             & 2 & $0.3< z_{\rm B} \leq 0.5$ & \num{10404223} & $0.59$ & $0.252$ & $-0.009\pm 0.007$ & $0.430$ & $0.396$ \\
                  & 3 & $0.5< z_{\rm B} \leq 0.7$ & \num{15508696} & $0.90$ & $0.276$ & $-0.010\pm 0.007$ & $0.546$ & $0.531$ \\
                  & 4 & $0.7< z_{\rm B} \leq 0.9$ & \num{9837460} & $0.64$ & $0.250$ & $0.008\pm 0.006$ & $0.744$ & $0.732$ \\
                  & 5 & $0.9< z_{\rm B} \leq 1.2$ & \num{8466542} & $0.59$ & $0.275$ & $0.006\pm 0.008$ & $0.909$ & $0.894$ \\
\hline
$T_{\rm B} > 3$   & 1 & $0.1< z_{\rm B} \leq 0.3$ & \num{7269125} & $0.42$ & $0.270$ & $-0.004\pm 0.008$ & $0.437$ & $0.244$ \\
(blue)            & 2 & $0.3< z_{\rm B} \leq 0.5$ & \num{12200673} & $0.75$ & $0.277$ & $-0.007\pm 0.006$ & $0.573$ & $0.431$ \\
                  & 3 & $0.5< z_{\rm B} \leq 0.7$ & \num{21116034} & $1.46$ & $0.292$ & $-0.002\pm 0.006$ & $0.791$ & $0.644$ \\
                  & 4 & $0.7< z_{\rm B} \leq 0.9$ & \num{12134896} & $0.92$ & $0.286$ & $0.026\pm 0.006$ & $0.914$ & $0.842$ \\
                  & 5 & $0.9< z_{\rm B} \leq 1.2$ & \num{10207426} & $0.87$ & $0.293$ & $0.036\pm 0.009$ & $1.081$ & $1.022$ \\

\hline
\end{tabular}
\tablefoot{The effective number density $n_{\rm eff}$ is calculated from Eq.~(1) of \citet{CFHT_2012MNRAS.427..146H}. The reported ellipticity dispersion is defined as $\sigma_{\epsilon,i}=(\sigma_{\epsilon 1}+\sigma_{\epsilon 2})/2$. The $m$-bias is defined in Eq.~(\ref{equ:shearbias}) and detailed in Sect.~\ref{Sec:cal_shear}. Reported uncertainties were computed from the dispersion of $50$ bootstrap samples. The mean and median of the redshift distributions were obtained from the DIR calibration, which is detailed in Sect.~\ref{Sec:cal_z}.}
\end{table*}

\section{Calibration of redshift distributions}\label{Sec:cal_z}

One of the most challenging tasks for a tomographic cosmic shear study is to estimate the source redshift distribution for each tomographic bin. These intrinsic redshift distributions vary with galaxy samples, so we need to calibrate the photo-$z$ estimates in the two sub-samples, separately. We followed the fiducial technique, dubbed DIR in \Htwo, for this task. This method directly estimates the underlying redshift distributions of a photometric sample using deep spectroscopic redshift (spec-$z$) catalogues that overlap with the photometric survey. We shortly discuss our implementation of this method in this section and refer interested readers back to the original papers for more details~\citep{DIR_2008MNRAS.390..118L,K450_2017MNRAS.465.1454H,KV450_2020AA...633A..69H}.

The DIR method requires that the calibration sample (the spec-$z$ sample) spans, at least sparsely, the full extent of the multi-band magnitude space covered by the target sample (the photo-$z$ sample) and that the mapping from magnitude space to redshift space is unique. Therefore, the coverage of the spec-$z$ sample is essential for the accuracy of this method. We here used the same set of spec-$z$ catalogues as used in the fiducial KV450 cosmic shear analysis. It includes the zCOSMOS survey~\citep{zCOSMOS_2009ApJS..184..218L}, the DEEP2 survey~\citep{DEEP2_2013ApJS..208....5N}, the VIMOS VLT Deep survey~\citep{VVDS_2013AA...559A..14L}, the GAMA-G15Deep survey~\citep{G15Deep_2018MNRAS.479.3746K} and a combined catalogue provided by ESO in the \emph{Chandra} Deep Field South area\footnote{\url{http://www.eso.org/sci/activities/garching/projects/goods/MasterSpectroscopy.html}}. These independent spec-$z$ surveys with different lines-of-sight and depths minimise shot noise and sample variance in the calibration sample.

Since the spec-$z$ catalogues cannot fully represent the photometric sample, one needs to weight spec-$z$ objects to ensure a suitable match between the spectroscopic and photometric distributions. The method, based on a \emph{k}th nearest neighbour (\emph{k}NN) approach, is detailed in Sect. 3 of \citet{K450_2017MNRAS.465.1454H}. Briefly, it assigns weights to the spec-$z$ objects by comparing the volume densities of the spec-$z$ and photometric objects in the nine-band magnitude space ($ugriZYJHK_s$). Therefore, KiDS+VIKING-like observations are required in the same areas as the aforementioned spec-$z$ surveys. \Htwo\ have built these photometric observations from multiple ways given the availability of specific data sets in those spec-$z$ survey fields. We adopted the same sample and split it with the same criterion as used for the main KV450 sample to build two representatives of our two sub-samples. 

\begin{figure*}
\includegraphics[width=\textwidth]{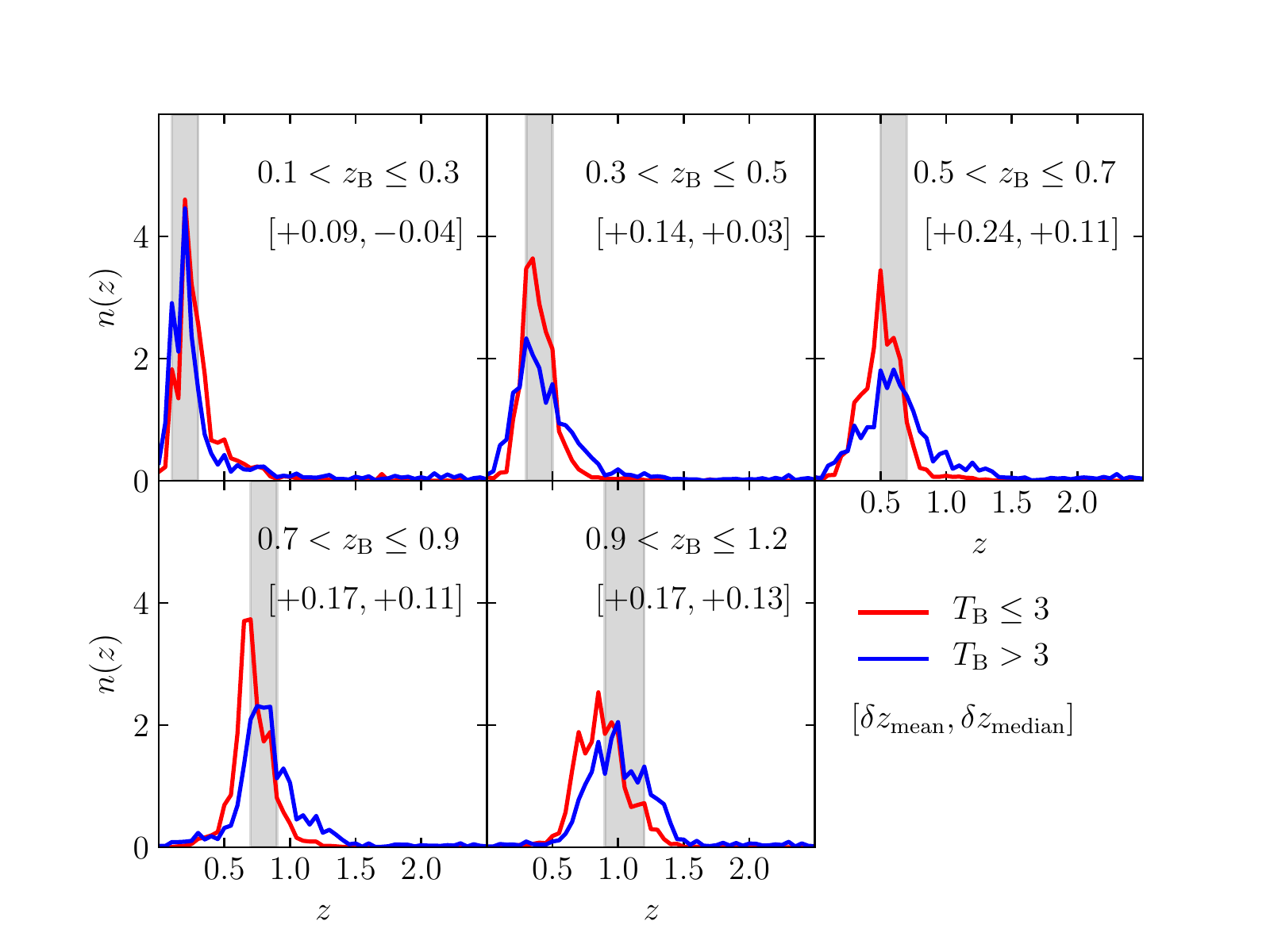}
\caption{\label{fig:z}Redshift distributions for the two sub-samples, estimated from DIR technique. Shaded regions correspond to photo-$z$ cuts for the tomographic binning. Mean and median differences were calculated as $\delta z_{\rm mean/median}=z_{\rm mean/median, blue}-z_{\rm mean/median, red}$.}
\end{figure*}

The resulting redshift distributions of the two sub-samples are shown in Fig.~\ref{fig:z}. Also presented are the mean and median differences between these two redshift distributions (see Table~\ref{Table:basic} for separate values). The importance of photo-$z$ calibration is demonstrated by the tails of the DIR redshift distributions compared to the ranges selected by the photo-$z$ cuts (shaded regions). These differences between the DIR results and photo-$z$ estimates are more significant in the red sub-sample, where an overall bias towards overestimating photo-$z$ is shown. This may seem counterintuitive at first given the discussion presented in Sect.~\ref{Sec:intro}, which states that young stars can wash out spectral features for photo-$z$ estimation resulting in larger errors in bluer galaxies. However, we stress that the red sub-sample defined in Sect.~\ref{Sec:data} is not `purely red', but also includes Sbc and Scd types (see Sect.~\ref{Sec:data}), which could worsen the photo-$z$ estimates. For our purposes, we are interested in the redshift difference between the two sub-samples. As can be seen, the differences are significant with the median differences as high as $\sim 0.13$ and the mean differences $\sim 0.24$ in certain bins. This level of difference will result in considerably different cosmic shear signals for the two sub-samples (see Sect.~\ref{Sec:shear}).

In practice, the DIR method is susceptible to various systematic effects, mainly induced by the incompleteness of the spec-$z$ sample, due to selection effects and sample variance in the different spectroscopic surveys that make up the spec-$z$ catalogue~(see \citealt{SOM_2020AA...637A.100W}, for an updated method that is more robust to such incompleteness). To account for these potential systematic effects, \Htwo\ introduced five nuisance parameters $\delta_{z_i}$ in their model to allow for linear shifts of the redshift distributions $n_i(z)\rightarrow n_i(z+\delta_{z_i})$ (see Table.~\ref{Table:initial}). Priors for these parameters are obtained using a spatial bootstrapping approach. In our consistency tests described below we focus on an extension of these nuisance parameters to the colour-split sub-samples (see Sect.~\ref{Sec:test}).

\section{Calibration of shape measurements}\label{Sec:cal_shear}

The shape measurements are susceptible to various biases due to the noise of galaxy images, the complexity of galaxy shapes, the selection effects and so on~(see Sect.~2 of \Kone, for a theoretical discussion). The weak lensing community have performed several blind challenges to test the performance of shape measurement pipelines~\citep{shearbias_2006MNRAS.368.1323H,shearbias_2007MNRAS.376...13M,shearbias_2010MNRAS.405.2044B,shearbias_2012MNRAS.423.3163K,shearbias_2015MNRAS.450.2963M}. These tests, based on simplified image simulations, are useful to understand common sources of shear bias, but cannot eliminate biases in a specific survey. In particular, differences in selection criteria between surveys affect the shear bias. These residual biases need to be calibrated with dedicated, tailor-made image simulations~\citep{shearbias_2015MNRAS.449..685H}. Following \citet{shearbias_2006MNRAS.368.1323H}, we quantify these residual biases using a linear parameterisation
\begin{equation}
\label{equ:shearbias}
g^{\rm obs}_i = (1+m_i)g^{\rm true}_i + c_i~,
\end{equation}
where $g^{\rm obs}_i$ and $g^{\rm true}_i$ are the observed and the true gravitational shears, respectively, with $i=1,2$ referring to the two different components. In practice, we found isotropy of $m$ results, that is $m_1\approx m_2$, so we simply adopt $m=(m_1+m_2)/2$.

\begin{figure*}
  \includegraphics[width=\textwidth]{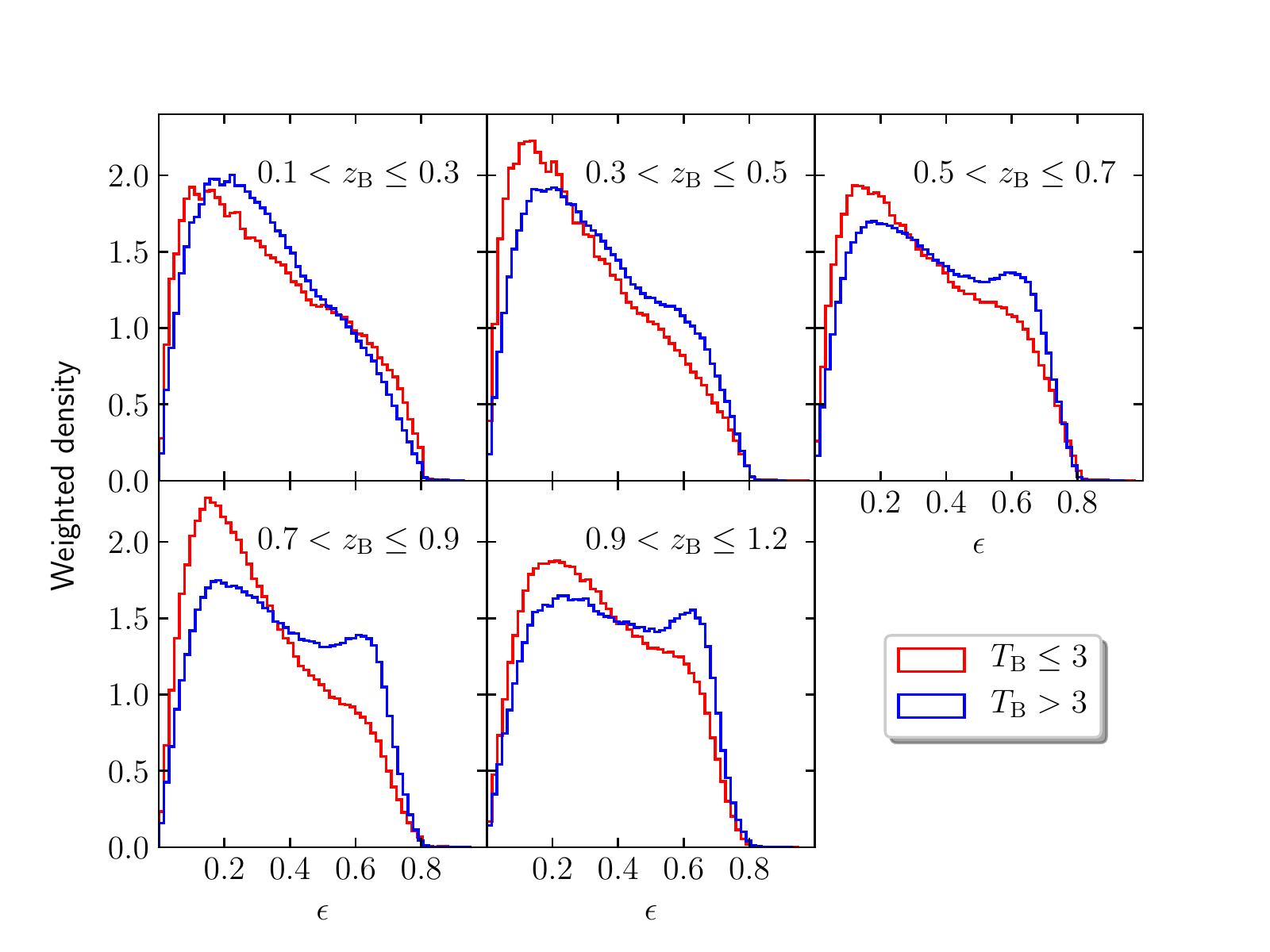}
    \caption{ \label{fig:e}Normalised \emph{lens}fit-weighted distributions of ellipticities of galaxies in the two sub-samples. The ellipticity is defined as $\epsilon=\sqrt{\epsilon_1^2+\epsilon_2^2}$. We note that the different distributions reflect different galaxy populations and indicate different shear biases in the two sub-samples.}
\end{figure*}

The two types of biases $m$ (the multiplicative bias) and $c$ (the additive bias or $c$-term) have different sources and properties. The former is usually determined from image simulations, whereas the latter can be inferred directly from the data. As \Kone\ show, shear biases depend not only on the selection function but also on the overall population of the galaxies. Therefore shear calibrations should be performed separately for samples containing different galaxy populations. This was the case for the different tomographic bins in the KV450 analysis and applies even more so to our split analysis. 

We therefore re-estimated multiplicative biases in the two sub-samples using the COllege simulations~(COSMOS-like lensing emulation of ground experiments, \Kone), which were also used in the current KV450 cosmic shear analysis. The main features of the COllege simulations are the observation-based input catalogue and the assignment of photometric redshifts. The input catalogue contains information on galaxy morphology and position from \emph{Hubble} Space Telescope observations~\citep{HST_2012ApJS..200....9G} of the COSMOS field~\citep{COSMOS_2007ApJS..172....1S}. The photometric redshifts of simulated galaxies are assigned by cross-matching the input catalogue to the KiDS catalogue. This setup ensures a high level of realism of the simulated catalogue and allows us to analyse the simulated data using the same pipelines as for the real data. \Kone\ have demonstrated that the simulated catalogue matches the full KV450 catalogue faithfully in all crucial properties including the galaxy shapes, sizes and positions. 

As expected, we found noticeable differences in the galaxy properties for the two sub-samples. We demonstrate one of these comparisons in Fig.~\ref{fig:e}, which compares the distributions of galaxy ellipticities. As already mentioned in Sect.~\ref{Sec:intro}, the ellipticity variance is one of the main sources of shape measurement biases~\citep[see also][]{shearbias_2014MNRAS.439.1909V} and therefore an indication of the variance of shear biases in the two sub-samples.

\begin{figure}
   \centering
   \includegraphics[width=\hsize]{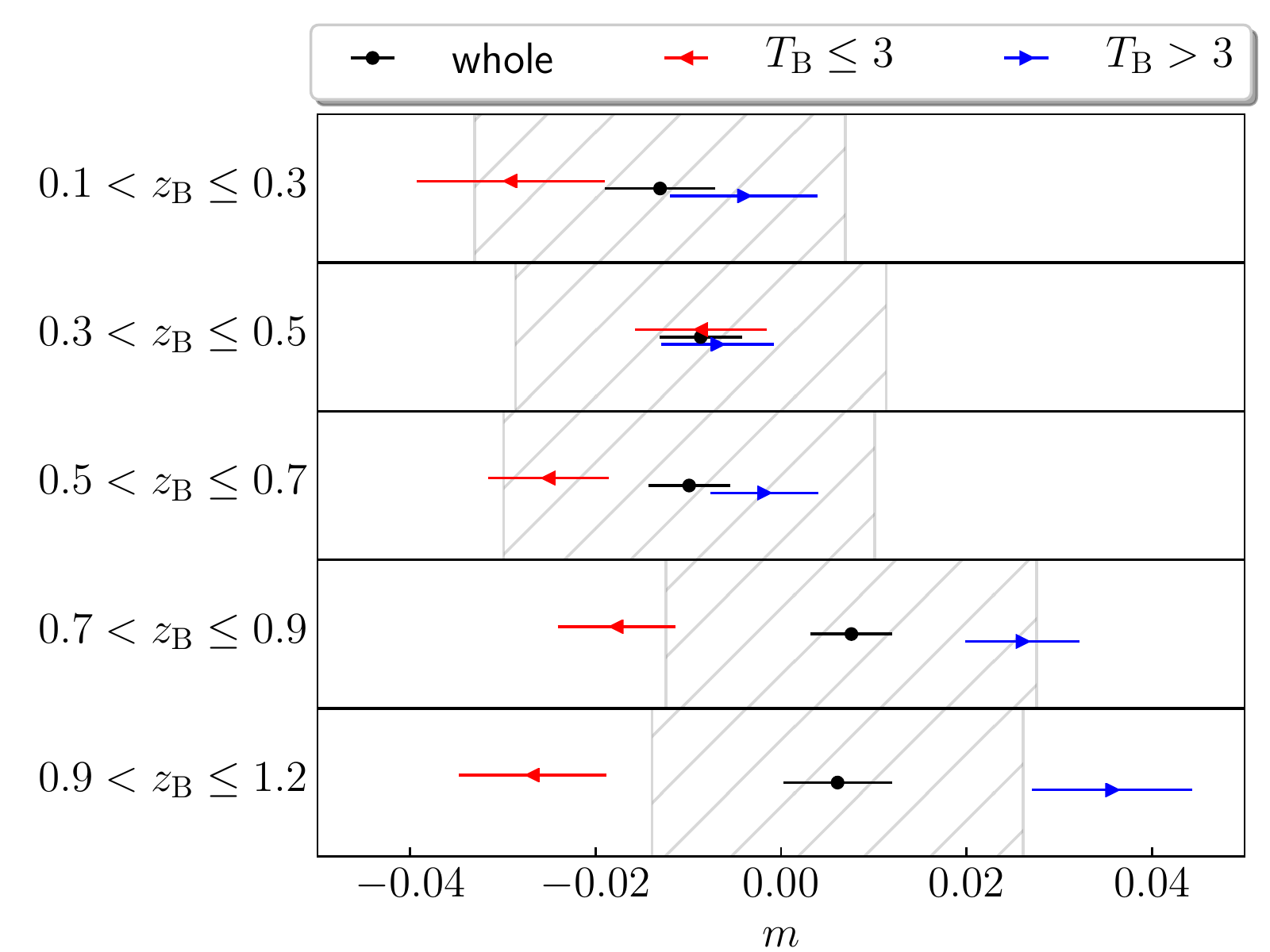}
      \caption{Multiplicative biases for the two sub-samples and the whole sample in each tomographic bin. Errors shown were estimated from bootstrapping. The hatched regions indicate the $0.02$ error budget adopted by \Htwo.}
         \label{fig:mbias}
\end{figure}
   
Our calibration approach is identical to that used in the fiducial KV450 cosmic shear analysis. It adopts a re-weighting scheme named as ``Method C'' in \citet{ImSim_2017MNRAS.467.1627F} to account for slight differences between the observations and the simulations. The $m$ value is reported per tomographic bin using a weighted average of individual galaxies belonging to the corresponding tomographic bin. We refer readers to Sect.~6 of \Kone\ for details.

We show our estimates of multiplicative biases for the two sub-samples in Fig.~\ref{fig:mbias}, compared with the results from the whole sample. The five sections from top to bottom correspond to the five tomographic bins from lower to higher redshifts. We noticed some significant differences in the $m$ values, especially for higher tomographic bins: these are mainly caused by the differences in the ellipticity distributions presented in Fig.~\ref{fig:e}. However, when considering the impact on the cosmic shear signals, the adjustments induced by these $m$-value differences are much smaller than those caused by the redshift differences as demonstrated in Fig.~\ref{fig:xipm}. We thus assumed that residual systematics from the shear calibration are secondary and focus our consistency tests on the redshift calibration.

The treatment of additive bias is sophisticated in the fiducial KV450 cosmic shear analysis~(see Sect.~4 of \Htwo, for details). Briefly, the treatment can be summarised as three aspects: First, the value of $c_i$ in each tomographic bin and in each patch is estimated by averaging over the measured galaxy ellipticities. These $c_i$ values are then subtracted from the galaxy ellipticities before the shear correlation functions are calculated (Eq.~\ref{eq:CF}). Second, a nuisance parameter $\delta_c$ is introduced into the model to account for a potential offset of the empirically determined $c_i$ values. The result from forward-modelling suggests that $\delta_c$ is very close to $0$ (see Table~\ref{Table:initial}). Third, a position-dependent additive bias pattern in the $\epsilon_1$ ellipticity component is introduced to account for an imperfection in the OmegaCAM detector chain. This pattern is publicly available as a supplementary file along with the main cosmic shear catalogues. Furthermore, another nuisance parameter $A_c$ is introduced to allow an overall scaling of this 2D pattern (see Table~\ref{Table:initial}).

We mainly followed this strategy for the additive bias calibration. We corrected the $c$-term per tomographic bin and per patch using the same empirical approach mentioned above. We also included the 2D $c$-term pattern in our models. But we abandoned the two nuisance parameters $\delta_c$ and $A_c$ from our model, as they do not have a significant impact on the fit.

\section{Shear signal}\label{Sec:shear}

The cosmic shear signal is encoded in the measured shapes of source galaxies as small coherent distortions. Therefore, proper statistical measures and models are required for a cosmic shear study. We detail these processes in this section. We first built the joint data vector for the two sub-samples with estimates of the shear correlation functions in Sect.~\ref{Sec:shear_measure} and then modelled it taking various astrophysical and cosmological effects into account in Sect.~\ref{Sec:shear_model}. The setup is based on the fiducial analysis of \Htwo\ but with slight adjustments to meet our test purpose.

\subsection{Statistical measures}\label{Sec:shear_measure}

The shear signal is captured by two-point shear correlation functions, which can be estimated from two tomographic bins $i$ and $j$ as
\begin{equation}
\label{eq:CF}
    \xi^{ij}_{\pm}(\theta)=\frac{\sum_{ab}w_aw_b\left[\epsilon_t^i(\bm{x}_a)\epsilon_t^j(\bm{y}_b)\pm\epsilon_{\times}^i(\bm{x}_a)\epsilon_{\times}^j(\bm{y}_b)\right]}{(1+m^i)(1+m^j)\sum_{ab}w_aw_b}~,
\end{equation}
where $\epsilon_{t,\times}$ are the tangential and cross ellipticities regarding the vector $\bm{x}_a-\bm{y}_b$ between a pair of galaxies $(a,b)$, and $w$ is the \emph{lens}fit weight. The summation runs over all galaxy pairs within an assigned spatial bin $\Delta\theta$ for each $\theta=|\bm{\theta}_b-\bm{\theta}_a|$. The multiplicative biases $m^{i}$ were obtained in Sect.~\ref{Sec:cal_shear} for each tomographic bin $i$.

We calculated Eq.~(\ref{eq:CF}) for the two sub-samples, separately, using the public \textsc{TREECORR} code\footnote{\url{https://github.com/rmjarvis/TreeCorr}}~\citep{TREECORR_2004MNRAS}. The spatial binning is identical to that used in \Htwo, that is, nine logarithmically spaced bins within the interval $[0.5', 300']$. We used the first seven bins for $\xi_{+}$, and the last six bins for $\xi_{-}$. These criteria are chosen to mitigate baryon feedback on small scales and the additive shear biases on large scales~(see \Htwo, for details). The joint data vector ($\xi^{\rm blue}_{\pm},~ \xi^{\rm red}_{\pm}$) we built through these measurements contains $(7+6)\times 15\times 2=390$ points.

\begin{figure*}
\includegraphics[width=\hsize]{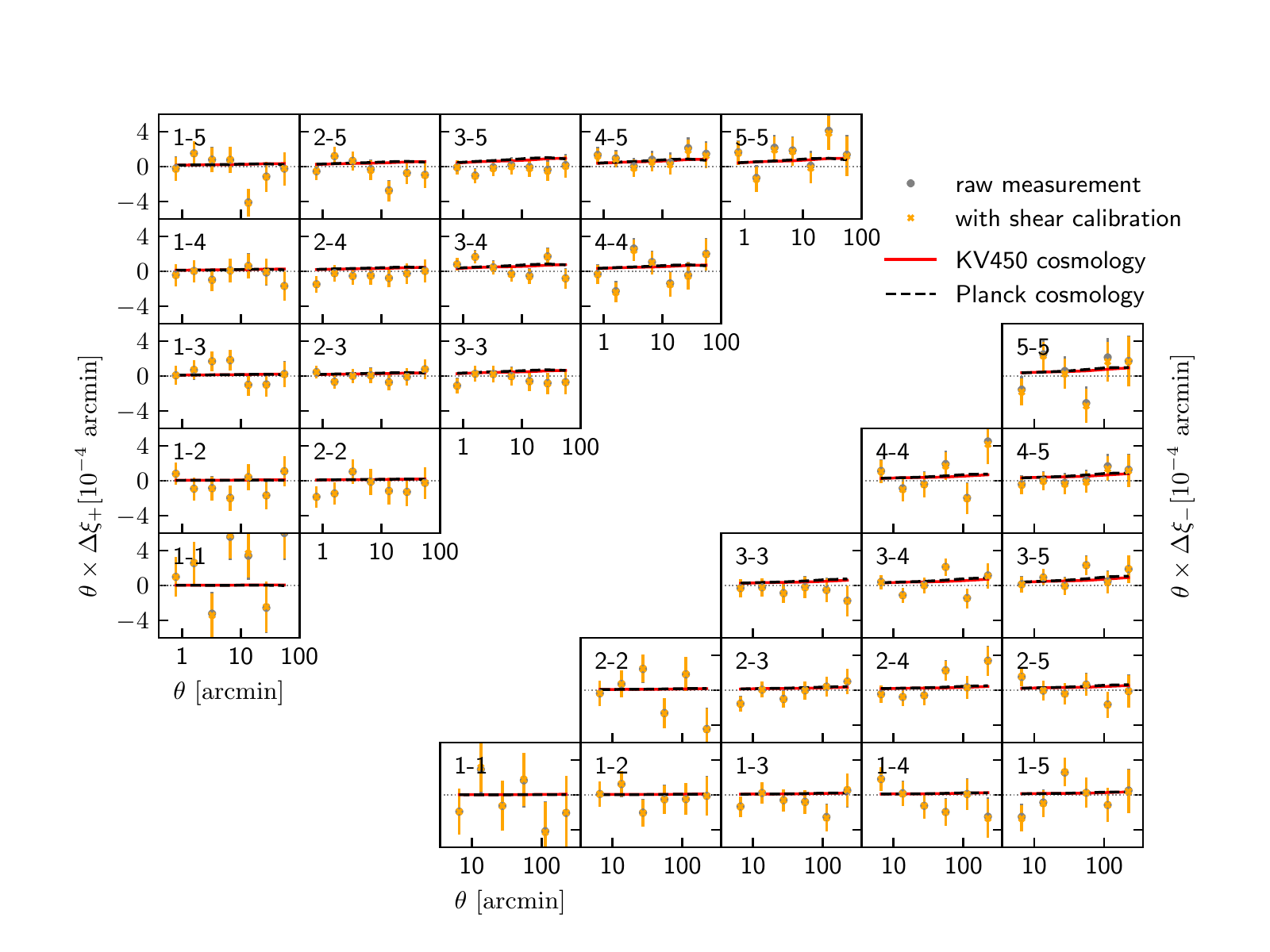}
\caption{\label{fig:xipm} Difference between two-point shear correlation functions from the two sub-samples ($\Delta\xi_{\pm} = \xi^{\rm blue}_{\pm}-\xi^{\rm red}_{\pm}$). The errors shown are defined as $\sigma_C=\sqrt{C_{\rm b,D}+C_{\rm r,D}-2C_{\rm br,D}}$, where the subscript `D' means the diagonal of a matrix, and the three unique parts of the whole covariance matrix are denoted as $C_{\rm b}$ for the blue sub-sample, $C_{\rm r}$ for the red sub-sample and $C_{\rm br}$ for their cross-covariance. We found these errors are close to the measurement errors reported by the TREECORR code ($\sigma_{\rm measure}/\sigma_C\gtrsim 0.8$), indicating that the diagonal elements of the covariance matrix are dominated by measurement noise. The overall agreement between the two sets of data vectors with and without the shear calibration (orange crosses vs.\ black dots) indicates the multiplicative bias has little effect in this study.}
\end{figure*}

We show our estimates of the data vector in Fig.~\ref{fig:xipm} with differences defined as $\Delta\xi_{\pm}=\xi^{\rm blue}_{\pm}-\xi^{\rm red}_{\pm}$. The errors shown were adopted from the analytical covariance matrix detailed in Sect.~\ref{Sec:cov}. Two series of data vectors correspond to the results with and without the multiplicative shear calibration. The difference is minor as expected given the overall small $m$ values (see Table~\ref{Table:basic}). Some non-zero trends are present in several bins, which are in principle caused by the different redshift distributions of these two sub-samples, as shown in Fig.~\ref{fig:z}. We detail how the redshift distributions can explain these measurements in the following section.

\subsection{Theoretical modelling}\label{Sec:shear_model}

The measured correlation functions $\xi^{ij}_{\pm}(\theta)$ are related to the lensing convergence power spectrum $P^{ij}_{\kappa}(\ell)$ through~\citep[see e.g.][]{review_2001PhR...340..291B}
\begin{equation}\label{eq:xiP}
    \xi^{ij}_{\pm}(\theta) =\frac{1}{2\pi}\int {\rm d}\ell~\ell P^{ij}_{\kappa}(\ell)J_{0/4}(\ell\theta)~,
\end{equation}
where $\ell$ is the angular wavenumber in the Fourier domain, and $J_{0/4}(\ell\theta)$ are Bessel functions of the first kind, with $J_0$ denoting the zeroth-order (for $\xi_{+}$) and $J_4$ the fourth-order (for $\xi_{-}$). Using the Kaiser-Limber approximation \citep{KL_1953ApJ...117..134L,KL_1992ApJ...388..272K,KL_1998ApJ...498...26K,KL_2008PhRvD..78l3506L}, $P^{ij}_{\kappa}(\ell)$ is in turn related to the physical matter power spectrum $P_{\delta}$, via
\begin{equation}\label{eq:PP}
    P_{\kappa}^{ij}(\ell)=\int^{\chi_{\rm H}}_0{\rm d}\chi~\frac{q_i(\chi)q_j(\chi)}{\left[f_{\rm K}(\chi)\right]^2}P_{\delta}\left(\frac{\ell+1/2}{f_{\rm K}(\chi)}, \chi\right)~,
\end{equation}
where $\chi$ and  $f_{\rm K}(\chi)$ are the comoving radial distance and the comoving angular distance, respectively. The upper limit of the integral $\chi_{\rm H}$ is the comoving horizon distance. The lensing efficiency $q_i(\chi)$ for tomographic bin $i$ is defined as
\begin{equation}\label{eq:qn}
    q_i(\chi) = \frac{3H_0^2\Omega_{\rm m}}{2c^2} \frac{f_{\rm K}(\chi)}{a(\chi)} \int^{\chi_{\rm H}}_{\chi}{\rm d}\chi'~n_i(\chi')\frac{f_{\rm K}(\chi'-\chi)}{f_{\rm K}(\chi')}~,
\end{equation}
which depends on the redshift distribution of galaxies $n_i(\chi){\rm d}\chi=n_i(z){\rm d}z$ along with other cosmological parameters. Therefore, different redshift distributions will cause a difference in shear signal between the two sub-samples.

We calculated the matter power spectrum using the Boltzmann-code \textsc{CLASS}~\citep{CLASS_2011JCAP...07..034B} with non-linear corrections from \textsc{HMCode}~\citep{hmcode_2016MNRAS.459.1468M}. Following \Htwo, we assumed a $\Lambda$CDM model with five primary cosmological parameters and one parameter for baryonic feedback processes on small scales. They are the densities of cold dark matter and baryons ($\Omega_{\rm CDM}$ and $\Omega_{\rm b}$), the amplitude and the index of the scalar power spectrum ($\ln(10^{10}A_{\rm s})$, $n_{\rm s}$), the scaled Hubble parameter ($h$), and the amplitude of the halo mass-concentration relation ($B$).

For the purposes of consistency tests, it is unnecessary to explore this whole cosmological parameter space, which is the same for the two sub-samples. Therefore, we fixed aforementioned cosmological parameters to two different sets of best-fit values from KV450~\citep{KV450_2020AA...633A..69H} and \emph{Planck}~\citep{Planck_2018arXiv180706209P} (see Table~\ref{Table:initial}). In this way, we can simplify our theoretical models while checking for potential cosmological dependence.

The last piece of information needed for modelling the observed correlation functions is the intrinsic alignment (IA) of galaxies \citep{IA_2015PhR...558....1T,IA_2015SSRv..193....1J}.  A common approach to make allowances for this effect is to add a ``non-linear linear'' IA model into the measured shear signal \citep{IA_2004PhRvD..70f3526H,IA_2007NJPh....9..444B}:
\begin{equation}\label{equ:xixi}
    \hat{\xi}_{\pm} = \xi_{\pm} +\xi^{\rm II}_{\pm} + \xi^{\rm GI}_{\pm}~,
\end{equation}
where $\hat{\xi}_{\pm}$ and $\xi_{\pm}$ correspond to the measured shear signal and the pure cosmic shear signal, respectively. The IA signals are added as $\xi^{\rm II}_{\pm}$ (`intrinsic-intrinsic' term between the intrinsic ellipticities of nearby galaxies) and $\xi^{\rm GI}_{\pm}$ (`gravitational-intrinsic' term between the intrinsic ellipticity of a foreground galaxy and the shear experienced by a background galaxy). These two IA terms can be calculated using the same formula shown in Eq.~(\ref{eq:xiP}) with power spectra
\begin{equation}\label{equ:PII}
    P^{ij}_{\rm II}(\ell) = \int_{0}^{\chi_{\rm H}}d\chi~F^2(z)\frac{n_i(\chi)n_j(\chi)}{\left[f_{\rm K}(\chi)\right]^2}P_{\delta}\left(\frac{\ell+1/2}{f_{\rm K}(\chi)}, \chi\right)~,
\end{equation}
\begin{equation}\label{equ:PGI}
    P^{ij}_{\rm GI}(\ell) = \int_{0}^{\chi_{\rm H}}d\chi~F(z)\frac{q_i(\chi)n_j(\chi)+q_j(\chi)n_i(\chi)}{\left[f_{\rm K}(\chi)\right]^2}P_{\delta}\left(\frac{\ell+1/2}{f_{\rm K}(\chi)}, \chi\right)~,
\end{equation}
where
\begin{equation}\label{equ:F}
    F(z) = -A_{\rm IA}C\rho_{\rm crit, 0}\frac{\Omega_{\rm m}}{D_{+}(z)}~.
\end{equation}
The normalisation constant is $C=5\times 10^{-14}h^{-1}M_{\odot}^{-1}{\rm Mpc}^3$, $\rho_{\rm crit, 0}$ is the critical density today, and the linear growth factor $D_{+}(z)$ is normalised to unity today. Following \Htwo, we ignored the redshift and luminosity dependence of IA and leave one nuisance parameter $A_{\rm IA}$ for IA effects (but see \citealt{IA_2020arXiv200302700F}). 

Now with all the information prepared, we can forward-model the shear correlation functions. For demonstration, we fixed all the model parameters and use the redshift distributions estimated in Sect.~\ref{Sec:cal_z} to predict the joint data vector of the two sub-samples. The results are shown in Fig.~\ref{fig:xipm}. Two different predictions come from two different sets of cosmological parameters: the red solid line from KV450 best-fit values and the black dashed line from \emph{Planck} best-fit values. All the other nuisance parameters are set to the best-fit KV450 results as shown in Table~\ref{Table:initial}. Even with this simple setting, the predicted results generally follow the trends seen from the data, demonstrating that the redshift difference is indeed the main cause for the different shear correlation functions in the two sub-samples. The other feature worth noting is the similarity between the two predictions from the two different sets of cosmological parameters. This implies that our test model is insensitive to the background cosmology. To quantify the goodness of fit and test the robustness of the pipelines, we need a more careful Bayesian analysis with proper test models and take correlations between measurements into account. 

%
\begin{table}
\caption{\label{Table:initial}Model parameters and their best-fit values from KV450 cosmic shear analysis~(\citealt{KV450_2020AA...633A..69H}) and \emph{Planck} CMB analysis~\citep{Planck_2018arXiv180706209P}.}
\centering
{\renewcommand{\arraystretch}{1.2}
\begin{tabular}{lrcl}
\hline\hline
Parameter & KV450 & {\it Planck} & Definition \\
\hline
$\Omega_{\rm CDM}h^2$ & $0.058$ & $0.120$ & CDM density today \\
$\Omega_{\rm b}h^2$ & $0.022$ & $0.022$ & Baryon density today \\
$\ln (10^{10}A_{\rm s})$ & $4.697$ & $3.045$ & Scalar spectrum amplitude \\
$n_{\rm s}$ & $1.128$ & $0.966$ & Scalar spectrum index \\
$h$ & $0.780$ & $0.673$ & Hubble parameter \\
\hline
$B$ & $2.189$ & - & Baryon feedback amplitude \\
$A_{\rm IA}$ & $0.494$ & - & IA amplitude \\
$\delta_c\times 10^{5}$ & $2.576$ & - & $c$-term offset \\
$A_c$ & $1.143$ & - & 2D $c$-term amplitude \\
$\delta_{z_1}$ & $-0.006$ & - & Bin 1 offset \\
$\delta_{z_2}$ & $0.001$ & - & Bin 2 offset \\
$\delta_{z_3}$ & $0.026$ & - & Bin 3 offset \\
$\delta_{z_4}$ & $-0.002$ & - & Bin 4 offset \\
$\delta_{z_5}$ & $0.003$ & - & Bin 5 offset \\
\hline
\end{tabular}
}
\tablefoot{The first five parameters are the standard cosmological parameters. Other parameters are nuisance parameters introduced by \citet{KV450_2020AA...633A..69H} to account for various effects associated with cosmic shear analysis. The KV450 best-fit values are extracted from the primary Monte Carlo Markov Chain, which is publicly available at \url{http://kids.strw.leidenuniv.nl/cosmicshear2018.php}. The \emph{Planck} best-fit values correspond to the TT,TE,EE+lowE+lensing results with the \texttt{Plik} likelihood (Table 1 of \citealt{Planck_2018arXiv180706209P}).}
\end{table}

\section{Consistency tests}\label{Sec:test}

Quantifying the internal consistency is not a trivial task given the correlations between measurements and the difficulty in comparing different models. On the one hand, neglecting intrinsic correlations between measurements can lead to untrustworthy conclusions. As demonstrated by \citet{consistency_2019MNRAS.484.3126K}, a lack of consideration of correlations can confuse residual systematics with the overall goodness of fit. On the other hand, null tests based on global summary statistics, such as Bayesian evidence, are practically difficult for high-dimensional models (see e.g. \citealt{review_2008ConPh..49...71T}). Moreover, different prior choices between hypotheses can complicate the interpretation of the final results~\citep{suspicious_2019PhRvD.100d3504H,suspicious_2019arXiv191007820L}. 

We address these issues in this section. We first built an analytical covariance matrix to account for all the correlations between measurements (Sect.~\ref{Sec:cov}). We then performed a Bayesian analysis with dedicated test parameters to quantify the potential discrepancy between measurements from the two sub-samples (Sect.~\ref{Sec:setup}). The conclusion is based on the posterior distributions of these test parameters. Through this approach, we can balance accuracy and simplicity in our model.

The modelling pipeline detailed below is publicly available\footnote{\url{https://github.com/lshuns/montepython_KV450}}. It is a modified version of the \textsc{MontePython} package~\citep{Monte_2013JCAP...02..001A,Monte_2018arXiv180407261B} with the \textsc{PyMultiNest} algorithm~\citep{pymultinest_2014AA...564A.125B}, which is a \textsc{python} wrapper of the nested sampling algorithm \textsc{MultiNest}~\citep{multinest_2009MNRAS.398.1601F}. The original \textsc{MontePython} package is adopted for the KV450 cosmological analysis in \Htwo\ and the consistency tests with a split of data vector~\citep{consistency_2019MNRAS.484.3126K}.

\subsection{Covariance matrix}\label{Sec:cov}

We estimated the covariance matrix for the joint data vector built in Sect.~\ref{Sec:shear_measure} using the analytical model developed in \citet{K450_2017MNRAS.465.1454H}, \Htwo\ and \citet{K1000method_2020arXiv200701844J}. The analytical approach is an improvement over the usual numerical or Jackknife approach with advantages in dealing with effects from modelling the noise and the finite survey areas. We here only briefly summarise the main features of this analytical recipe and refer interested readers to Sect.~5 of \citet{K450_2017MNRAS.465.1454H} and \citet{K1000method_2020arXiv200701844J} for details.

The analytical model comprises three terms: a Gaussian term associated with sample variance and shape noise, a non-Gaussian term from in-survey modes, and a third term, which is also non-Gaussian, from super-survey modes (known as super-sample covariance; SSC). The first, Gaussian term is estimated following \citet{covariance_2008AA...477...43J}, with a transfer function from \citet{transfer_1998ApJ...496..605E} and the non-linear corrections from \citet{nonlinear_2012ApJ...761..152T}. The source information used is listed in Table~\ref{Table:basic}; these are the effective galaxy number density ($n_{\rm eff}$) and the weighted ellipticity dispersion ($\sigma_{\epsilon, i}$). The second, non-Gaussian term is calculated using the formalism from \citet{covariance_2013PhRvD..87l3504T} with the halo mass function and halo bias from \cite{halo_2010ApJ...724..878T}. The halo profile is described using a Fourier-transform version~\citep{NFW_2001ApJ...546...20S} of the NFW model~\citep{NFW_1996ApJ...462..563N}, with the concentration-mass relation from \citet{concentration_2008MNRAS.390L..64D}. The final, SSC term is again modelled using the formalism from \citet{covariance_2013PhRvD..87l3504T}, and the survey footprint is modelled with a \textsc{HEALPix} map~\citep{healpix_2005ApJ...622..759G}.

The shear calibration presented in Sect.~\ref{Sec:cal_shear} also suffers from uncertainties. We adopted a systematic uncertainty $\sigma_{m}=0.02$ for the multiplicative biases as estimated by \Kone\ and used in \Htwo\ and~\cite{KV450SOM_2020arXiv200504207W} and propagated it into the covariance matrix through $C^{\rm cal}_{ij}=4\xi^{\rm T}_{i}\xi^{\rm T}_{j}\sigma^2_{m}+C_{ij}$, where $\xi^{\rm T}$ is the joint data vector predicted using the KV450 best-fit values and the DIR redshift distributions (see Sect.~\ref{Sec:cal_z}). We ignored the error of the additive biases due to its negligible effect~\citep[see Appendix D4 of][for a detailed discussion]{K450_2017MNRAS.465.1454H}. 

\begin{figure}
   \centering
   \includegraphics[width=\hsize]{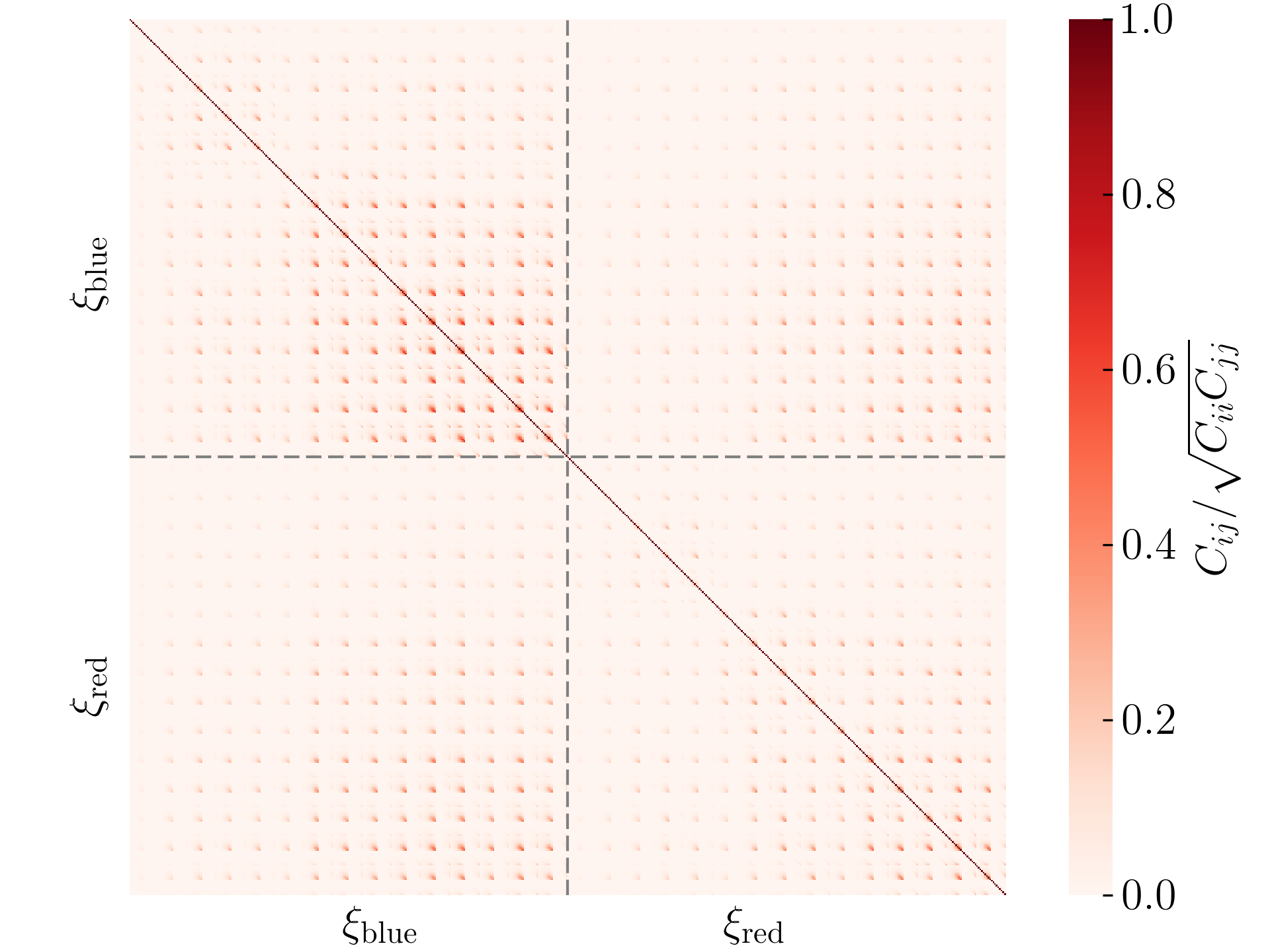}
      \caption{Analytical correlation matrix for the joint data vector. The covariance $C_{ij}$ is normalised using the diagonal $\sqrt{C_{ii}C_{jj}}$ to show the correlation matrix.}
         \label{fig:cov}
   \end{figure}

We show the final correlation matrix for the joint data vector in Fig.~\ref{fig:cov}. Non-negligible contributions from off-diagonal regions are easily noticed, indicating the non-trivial correlations between the measurements both within individual sub-samples and across the two sub-samples. The importance of the potential correlations between (two) parts of a split was already highlighted in \citet{consistency_2019MNRAS.484.3126K}, but here we confirmed it more directly. By including the full covariance matrix into our consistency tests, we naturally took all the data correlations into account.

We inspected the relative contributions of the Gaussian and non-Gaussian terms to the full covariance matrix. We found that the Gaussian term generally dominates over the non-Gaussian term in the diagonal parts, but the latter contributes more in the off-diagonal regions. This general behaviour is more clearly demonstrated in \citet{K1000method_2020arXiv200701844J}. Since our test model is most sensitive to the difference $\Delta\xi$ between the two sub-samples, we constructed the covariance matrix of $\Delta\xi$ as $C_{\Delta}=C_{\rm blue}+C_{\rm red}-2C_{\rm cross}$, and compared it to the covariance matrices of the single data vectors ($\xi_{\rm blue}$ or $\xi_{\rm red}$). We found that the non-Gaussian contributions are significantly suppressed in $C_{\Delta}$ with an overall reduction of $\lesssim 75\%$ compared to $C_{\rm blue}$. The Gaussian contributions are also slightly suppressed, mainly in the off-diagonal regions. The cancellation of sample variance can explain both suppressions in the covariance matrix $C_{\Delta}$. Therefore, we verified that our test model is robust against uncertainties in the sample variance and changes in the cosmological parameters.

\subsection{Test setup}\label{Sec:setup}

With the covariance matrix prepared, we can now explore the parameter space with a Bayesian analysis. Our primary objective is to check if a common set of nuisance parameters is sufficient to capture residual systematics in the two sub-samples. For this purpose, we fixed all the cosmological parameters, which in principle share the same values in the two sub-samples. We verified this cosmology insensitivity assumption by running an additional chain with free cosmological parameters. The results are consistent with the fixed-cosmology settings, and we hardly observe any degeneracy between cosmological parameters and test parameters. In what follows, we therefore fixed the cosmology parameters to simplify the likelihood function and avoid unnecessary exploration of the high-dimensional parameter space. To account for any potential residual effects from an `incorrect' choice of cosmological parameters, we ran two setups with cosmological parameters from the KV450 cosmic shear results and from the \emph{Planck} CMB results (see Table~\ref{Table:initial}).

We built our test model $\mathcal{H}_1$ by introducing six test parameters besides the nuisance parameters used in \Htwo: a shift in IA amplitude $A_{\rm IA,s}$ and shifts in redshift offsets $\delta_{z_i,{\rm s}}$.  We implemented them in the two sub-samples as
\begin{equation}\label{eq:test}
X_{\rm blue/red} = X \pm X_{\rm s}~,
\end{equation}
where $X$ represents the $A_{\rm IA}$ or $\delta_{z_i}$ parameters, whereas $X_{\rm s}$ denotes corresponding test parameters. The plus sign is applied to the blue sub-sample, and the minus sign is for the red sub-samples. While a difference in the IA signal is expected, the differences in redshift offsets should vanish if the calibration pipeline is robust against sample-related systematics. Any non-vanishing values of $\delta_{z_i,{\rm s}}$ imply residual systematics that cannot be adequately captured by the common nuisance parameters. We therefore based our result mainly on the posterior distributions of these test parameters. In addition, we set up a base model $\mathcal{H}_0$ for a control purpose, where we adopted the same set of nuisance parameters as in \Htwo\ to model the joint data vector built from our two sub-samples. It includes six free nuisance parameters: the amplitude of the IA signal $A_{\rm IA}$ (see Sect.~\ref{Sec:shear_model}) and the redshift offset $\delta_{z_i}$ for each tomographic bin $i$ (see Sect.~\ref{Sec:cal_z}). This is a stronger assumption than what is required by the data consistency, since the IA signal, which depends on the galaxy population, is not expected to be the same for the two sub-samples.

Prior distributions for all the free parameters are listed in Table~\ref{Table:final}. The common nuisance parameters adopt priors from \Htwo, where $A_{\rm IA}$ has a wide flat prior, whereas $\delta_{z_i}$ have Gaussian priors with variance determined from a spatial bootstrapping approach during the redshift calibration (see Sect.~3.2 of \Htwo). The six new test parameters in the test model $\mathcal{H}_1$ use wide and uninformative priors. As will be shown in Sect.~\ref{Sec:resul}, these prior choices incorporate prior knowledge of redshift uncertainties into the common nuisance parameters and meanwhile allow for a thorough exploration of the test parameters. We stress that the main goal of our test is to evaluate the sufficiency of the KV450 nuisance parameters in capturing residual systematics.

Since we do not rely on the Bayesian evidence to diagnose tensions, our test method is free from the ``suspiciousness’’ problem linked to common model-selection methods~\citep{suspicious_2019arXiv191007820L}; in this respect, our test approach is analogous to the second tier of the Bayesian consistency tests proposed by \citet{consistency_2019MNRAS.484.3126K}. However, instead of duplicating the cosmological parameters and drawing conclusions based on the posterior distributions of cosmological parameter differences, we focus on the nuisance parameters, especially those linked to the redshift calibration. The other essential difference is that we performed a colour-based split of the source galaxies and re-did measurements and calibrations for the sub-samples, whereas \citet{consistency_2019MNRAS.484.3126K} based their comparison on a split of the measured correlation functions. Therefore, our method is more sensitive to possible inconsistencies within the source samples, whereas their approach is a more global test of residual systematics and the impact on the final cosmological results. In this sense, our test serves as a complementary check of the pipeline robustness to theirs.

\section{Results}\label{Sec:resul}

\begin{figure}
   \centering
   \includegraphics[width=\hsize]{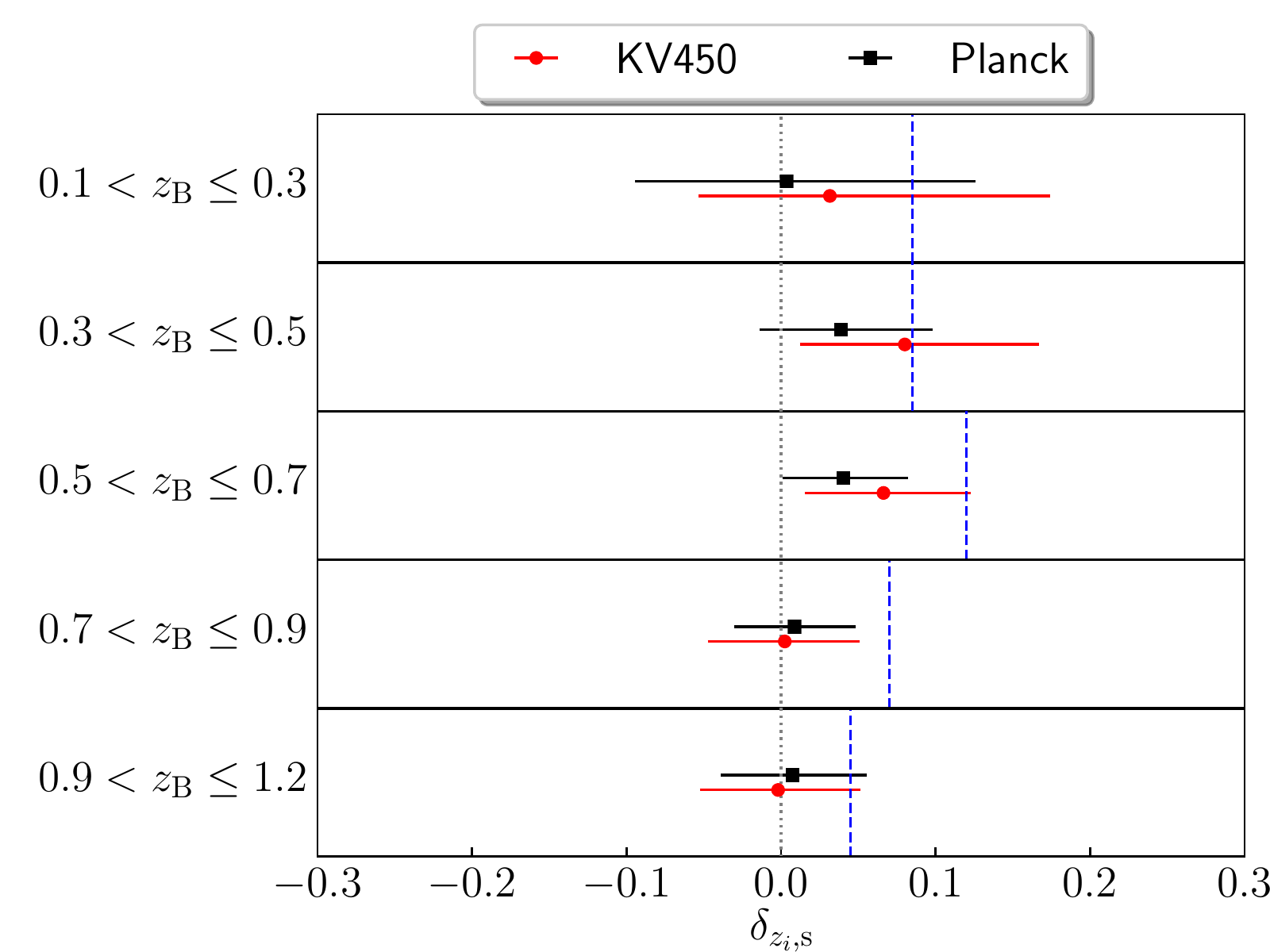}
      \caption{Constraints on $\delta_{z_i,{\rm s}}$ per tomographic bin for the $\mathcal{H}_1$ model. Errors shown correspond to the $68\%$ credible intervals from the MCMC run. For comparison, the vertical blue lines show the half of the mean differences between the reconstructed DIR redshift distributions of the two sub-samples (see Fig.~\ref{fig:z}).}
         \label{fig:Dzs}
   \end{figure}

The main results from our consistency tests are shown in Fig.~\ref{fig:Dzs}. These are the marginal posterior constraints of the five test parameters $\delta_{z_i, {\rm s}}$ introduced in Sect.~\ref{Sec:setup}. The five sections in the plot correspond to the five tomographic bins. The two sets of values are from the two sets of cosmological parameters we employed: the KV450 best-fit cosmology (red lines) and the \emph{Planck} best-fit cosmology (black lines). Both sets of results agree with each other, further confirming that our test model is insensitive to the choice of cosmological parameters. As can be seen, all values are consistent with zero within $\sim 1.5\sigma$, indicating that the KV450 calibration pipelines are correcting these sample-related systematics, and introducing more nuisance parameters is unnecessary for the current analysis. 

The two tomographic bins with slight non-vanishing differences are the second bin ($\sim 1.2\sigma$) and the third bin ($\sim 1.3\sigma$). Interpreting this level of difference is complex, given the statistical power of the current data. We reiterate that the $\delta_{z_i,{\rm s}}$ parameters we constrained here refer to the shifts of the redshift offsets in the two sub-samples. These are expected to be larger than the mean redshift offsets ($\delta_{z_i}$), given the substantial redshift differences between the two sub-samples and the width of the DIR redshift distributions (see Fig.~\ref{fig:z}). As seen from Table~\ref{Table:final}, all $\delta_{z_i,{\rm s}}$ values are smaller than the width of the underlying redshift distributions and are close to zero within the uncertainties. This reflects the overall accuracy of the DIR redshift distributions.

Table~\ref{Table:final} lists the posterior results for all free parameters and the best-fit $\chi^2$-values for all models. We do not base our conclusion on the $\chi^2$-test, because the dimensionality is not directly specified by the number of free parameters in a complex Bayesian model~\citep[see e.g][]{dimension_2019PhRvD.100b3512H}. Nevertheless, a simple comparison of the best-fit $\chi^2$ values with the number of free parameters taken into account suggests that the test model $\mathcal{H}_1$ is indistinguishable from the control model $\mathcal{H}_0$. This lends some more credit to our previous conclusion on the adequacy of current nuisance parameters in dealing with residual systematics.

\begin{figure*}
\includegraphics[width=\hsize]{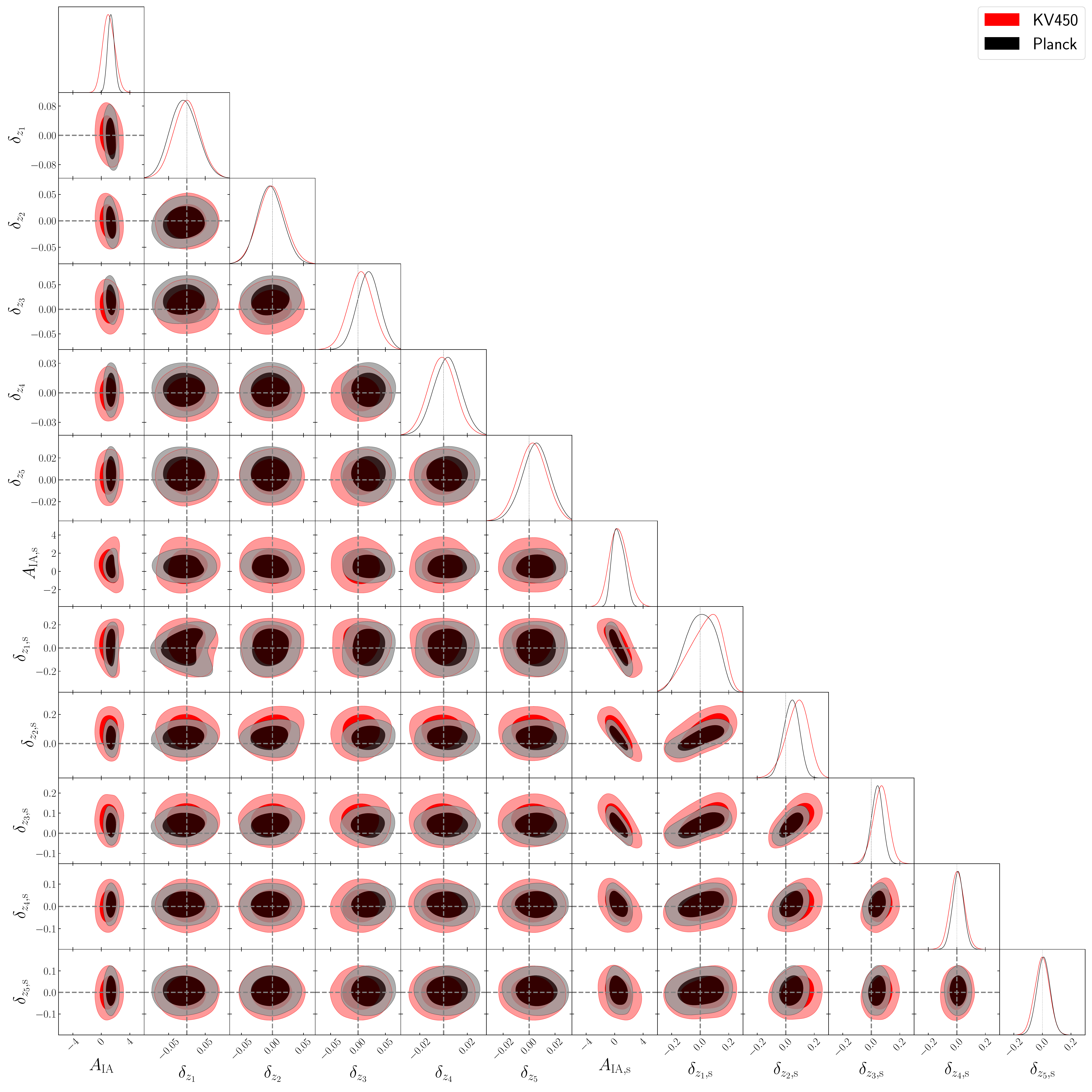}
\caption{\label{fig:H1} Contour plots of the $68\%$ and $95\%$ credible regions for all the free parameters in $\mathcal{H}_1$ model. Plotting ranges are the same as the prior ranges. Dashed lines indicate zero values in the ideal case. Two different colours correspond to the two sets of results from KV450 and \emph{Planck} cosmological values. The slight degeneracy between $\delta_{z_i,{\rm s}}$ in the low redshift bins is an effect from the high degeneracy between $A_{\rm IA,s}$ and $\delta_{z_i,{\rm s}}$. It vanishes in the test setting $\mathcal{T}_1$, where $A_{\rm IA,s}$ is fixed to zero.}
\end{figure*}

Figure~\ref{fig:H1} presents the contour plot for the test model. An interesting feature we note is the high degeneracy between $A_{\rm IA,s}$ and $\delta_{z_i, {\rm s}}$ in the low redshift bins (see Fig.~\ref{fig:H1}). This incurs most of the ambiguities in the test parameters. The entanglement between the IA signal and the redshift uncertainties is also noticed in \citet{KV450SOM_2020arXiv200504207W}, where a revised redshift calibration of the KV450 data results in a vanishing IA amplitude. Our finding affirms the difficulty in interpreting the apparent IA signal. We conducted an extreme test where we fixed $\delta_{z_i, {\rm s}}=0$ in the test model $\mathcal{H}_1$. It led to a large positive $A_{\rm IA,s}$ value, suggesting $A_{\rm IA, blue}>A_{\rm IA, red}$. This is inconsistent with dedicated IA studies (see \citealt{IA_2015SSRv..193....1J}, for a review), implying that IA parameters can disguise problems with the redshift estimates. Therefore, we should be careful to interpret the IA parameters. To check the impact of the IA parameters in our test model, we ran one more test $\mathcal{T}_1$, in which $A_{\rm IA,s}$ was fixed to zero. This maximises the shifts of the redshift offsets by ignoring the IA difference in the two sub-samples. Even in this conservative estimate, the shifts are $\lesssim 2.1\sigma$ for all redshift bins, with the highest values again seen in the third bin~(see Table.~\ref{Table:final}).

%
\begin{table*}
\caption{\label{Table:final}Priors and posterior results for all models.}
\centering
{\renewcommand{\arraystretch}{1.5}
\begin{tabular}{lcrrrrrr}
\hline\hline
Parameter & Prior & \multicolumn{3}{c}{KV450} & \multicolumn{3}{c}{Planck} \\
\cline{3-8}
 & & $\mathcal{H}_0$ & $\mathcal{H}_1$ & $\mathcal{T}_1$ & $\mathcal{H}_0$ & $\mathcal{H}_1$ & $\mathcal{T}_1$ \\
\hline
$A_{\rm IA}$ & $[-6, 6]$ & $1.442_{-0.898}^{+0.826}$ & $1.049_{-0.871}^{+0.818}$ & $0.976_{-0.804}^{+0.776}$ & $1.741_{-0.533}^{+0.507}$ & $1.358_{-0.495}^{+0.463}$ & $1.340_{-0.476}^{+0.466}$\\
$\delta_{z_1}$ & $0.000\pm 0.039$ & $-0.012_{-0.037}^{+0.037}$ & $-0.000_{-0.038}^{+0.035}$ & $0.001_{-0.037}^{+0.035}$ & $-0.037_{-0.036}^{+0.028}$ & $-0.008_{-0.040}^{+0.036}$ & $-0.005_{-0.039}^{+0.038}$ \\
$\delta_{z_2}$ & $0.000\pm 0.023$ & $-0.006_{-0.023}^{+0.019}$ & $-0.001_{-0.021}^{+0.022}$ & $-0.000_{-0.022}^{+0.021}$ & $-0.011_{-0.019}^{+0.019}$ & $-0.003_{-0.022}^{+0.020}$ & $-0.002_{-0.019}^{+0.021}$ \\
$\delta_{z_3}$ & $0.000\pm 0.026$ & $0.009_{-0.022}^{+0.023}$ & $0.006_{-0.023}^{+0.022}$ & $0.006_{-0.026}^{+0.022}$ & $0.020_{-0.018}^{+0.020}$ & $0.019_{-0.021}^{+0.020}$ & $0.021_{-0.020}^{+0.020}$ \\
$\delta_{z_4}$ & $0.000\pm 0.012$ & $-0.002_{-0.011}^{+0.012}$ & $-0.001_{-0.011}^{+0.012}$ & $-0.002_{-0.012}^{+0.012}$ & $0.003_{-0.012}^{+0.011}$ & $0.003_{-0.012}^{+0.012}$ & $0.003_{-0.013}^{+0.012}$ \\
$\delta_{z_5}$ & $0.000\pm 0.011$ & $0.002_{-0.011}^{+0.011}$ & $0.003_{-0.010}^{+0.012}$ & $0.002_{-0.011}^{+0.011}$ & $0.006_{-0.011}^{+0.012}$ & $0.005_{-0.010}^{+0.011}$ & $0.006_{-0.011}^{+0.011}$ \\
\hline
$A_{\rm IA,s}$ & $[-6, 6]$ & - & $0.571_{-1.337}^{+1.178}$ & - & - & $0.536_{-0.967}^{+0.793}$ & - \\
$\delta_{z_1,{\rm s}}$ & $[-0.3, 0.3]$ & - & $0.032_{-0.085}^{+0.142}$ & $0.079_{-0.069}^{+0.076}$ & - & $0.004_{-0.098}^{+0.122}$ & $0.072_{-0.066}^{+0.057}$ \\
$\delta_{z_2,{\rm s}}$ & $[-0.3, 0.3]$ & - & $0.080_{-0.068}^{+0.087}$ & $0.116_{-0.055}^{+0.048}$ & -  & $0.039_{-0.053}^{+0.059}$ & $0.069_{-0.033}^{+0.032}$\\
$\delta_{z_3,{\rm s}}$ & $[-0.3, 0.3]$ & - & $0.066_{-0.051}^{+0.057}$ & $0.087_{-0.041}^{+0.037}$ & -  & $0.040_{-0.039}^{+0.042}$ & $0.060_{-0.030}^{+0.027}$\\
$\delta_{z_4,{\rm s}}$ & $[-0.3, 0.3]$ & - & $0.002_{-0.050}^{+0.048}$ & $0.014_{-0.045}^{+0.044}$ & -  & $0.009_{-0.039}^{+0.039}$ & $0.019_{-0.037}^{+0.037}$\\
$\delta_{z_5,{\rm s}}$ & $[-0.3, 0.3]$ & - & $-0.002_{-0.051}^{+0.053}$ & $0.005_{-0.050}^{+0.051}$ & -  & $0.008_{-0.046}^{+0.048}$ & $0.015_{-0.046}^{+0.046}$\\
\hline
$N_{\rm data}$ & - & $390$ & $390$ & $390$ & $390$ & $390$ & $390$\\
$N_{\rm para}$ & - & $6$ & $12$ & $11$ & $6$ & $12$ & $11$\\
$\chi^2$ & - & $366.8$ & $356.4$ & $356.1$ & $357.5$ & $364.5$ & $364.4$\\
\hline
\end{tabular}
}
\tablefoot{The first six parameters are common nuisance parameters to account for overall IA amplitude and redshift offsets. The following are six test parameters introduced to account for potential differences of aforementioned parameters between the two sub-samples (see Eq.~\ref{eq:test}). Priors shown in brackets are top-hat ranges whereas values with errors indicate Gaussian distributions. Results are the mean values of the posterior whereas the $\chi^2$ corresponds to the maximum likelihood. Two sets of results were derived, fixing cosmological parameters to either the KV450 or the \emph{Planck} values (see Table.~\ref{Table:initial}). The test model $\mathcal{H}_1$ contains $12$ free parameters. The `control model' $\mathcal{H}_0$ ignores parameter differences between the two sub-samples and only includes $6$ common parameters. The test setting $\mathcal{T}_1$ ignores the difference of IA signals in the two sub-samples. }
\end{table*}
%

\section{Summary and discussion}\label{Sec:discussion}

We presented an internal-consistency test to the KV450 cosmic shear analysis with a colour-based split of source galaxies, resulting in two statistically comparable sub-samples containing noticeably different galaxy populations (see Figs.~\ref{fig:TB}, \ref{fig:z} and \ref{fig:e}). We performed the same measurements and calibrations to these two sub-samples and assessed changes in the two-point correlation functions because of known differences in the redshift distributions and the multiplicative biases (see Fig.~\ref{fig:xipm}). By fixing cosmological parameters, we examined the internal consistency of the observational nuisance parameters, specifically those for the redshift distributions, using a Bayesian analysis with dedicated test parameters. We observed a degeneracy between the redshift uncertainties and the inferred IA amplitude for low redshift bins, but we found no evidence of internal inconsistency in the KV450 data, verifying that the current strategy of linearly shifting redshift distributions with a common set of nuisance parameters is adequate for capturing residual systematics in the redshift calibration.

The internal-consistency test we proposed is robust against the uncertainties of the background cosmology and cosmic variance. It can be implemented in future cosmic shear surveys before any cosmological inference is made. This weak sensitivity to cosmology is shared with the existing ``shear-ratio’’ test~\citep{shearratio_2003PhRvL..91n1302J,shearratio_2016AA...592L...6S,shearratio_2019AA...623A..94U}, which has already been applied to check the accuracy of redshift distributions in current cosmic shear surveys~(\citealt{shearratio_2012MNRAS.427..146H}; \Htwo; \citealt{K1000cata_2020arXiv200701845G}). The ``shear-ratio’’ test is a cross-correlation approach based on the galaxy-galaxy lensing signals of two or more source samples at different redshift bins. Therefore, the two tests are sensitive to different systematics, making them complementary.

Although our discussion concentrated on the redshift calibration, we found that the test also relies on our assumptions regarding the IA signals (see Fig.~\ref{fig:H1}). Without a thorough exploration of IA models, our test can already pick up the degeneracy between the IA signals and the redshift uncertainties, which has been implied in previous studies~(see Sect.~6.6 of \citealt{K450_2017MNRAS.465.1454H}). Recently, \citet{IA_2019MNRAS.489.5453S} performed an analogous split-based analysis to the DES data. They focus on the IA signal and cosmological parameters and marginalise over observational nuisance parameters. This is different from what we explored here, but connected to our test through the IA signals, which were examined in both tests. They provided better constraints on the IA signals in sub-samples using a variety of IA models. We can perform analogous improvements to our test model to learn more about the IA signals and their correlation to other nuisance parameters in future cosmic shear data.


\begin{acknowledgements}
    We thank Shahab Joudaki for carefully reading the manuscript and providing useful comments. SSL is supported by NOVA, the Netherlands Research School for Astronomy. KK acknowledges support by the Alexander von Humboldt Foundation. HHo acknowledges support from Vici grant 639.043.512, financed by the Netherlands Organisation for Scientific Research (NWO). HHi is supported by a Heisenberg grant of the Deutsche Forschungsgemeinschaft (Hi 1495/5-1) as well as an ERC Consolidator Grant (No. 770935). This work is based on observations made with ESO Telescopes at the La Silla Paranal Observatory under programme IDs 100.A-0613, 102.A-0047, 179.A-2004, 177.A-3016, 177.A-3017, 177.A-3018, 298.A-5015, and on data products produced by the KiDS consortium.    

\end{acknowledgements}

{\small \textit{Author Contributions:} All authors contributed to the development and writing of this paper. The authorship list is given in two groups: the lead authors (SSL, KK, HHo), followed by an alphabetical group of key contributors to both the scientific analysis and the data products.}

%
  \bibliographystyle{aa} 
  \bibliography{reference} 
%

\end{document}